\documentclass{JHEP3}
\pdfoutput=1

\JHEPspecialurl{http://jhep.sissa.it/JOURNAL/JHEP3.tar.gz}

\usepackage{graphicx}
\usepackage[tight]{subfigure}
\usepackage{amsmath}

\newcommand\fverb{\setbox\pippobox=\hbox\bgroup\verb}
\newcommand\fverbdo{\egroup\medskip\noindent%
			\fbox{\unhbox\pippobox}\ }
\newcommand\fverbit{\egroup\item[\fbox{\unhbox\pippobox}]}
\newbox\pippobox

\title{Constrained potential method for false vacuum decays}

\author{Jae-hyeon~Park\\
  Deutsches Elektronen-Synchrotron DESY,
  Notkestra{\ss}e 85, 22607 Hamburg, Germany\\
  \email{jae-hyeon.park@desy.de}}

\preprint{
DESY 10--213}

\abstract{%
A procedure is reported for numerical analysis of false vacuum transition
in a model with multiple scalar fields.
It is a refined version of the approach by Konstandin and Huber.
The alteration makes it possible to tackle a class of problems
that was difficult or unsolvable with the original method, i.e.\
those with a distant or nonexistent true vacuum.
An example with an unbounded-from-below direction is presented.}

\newcommand{\wt}{\widetilde}

\newcommand{\SE}{S}
\newcommand{\SEx}[1]{\wt{S}_{#1}}
\newcommand{\SEalpha}{\SEx{\alpha}}

\newcommand{\phibounce}{\overline{\phi}}
\newcommand{\phib}{\phi_b}
\newcommand{\phie}{\phi_e}
\newcommand{\phieinit}{\phie}
\newcommand{\phifv}{\phi_+}
\newcommand{\phitv}{\phi_-}
\newcommand{\offset}{\Delta \rho}

\newcommand{\x}{\phi_2}
\newcommand{\y}{\phi_1}
\newcommand{\msqx}{m^2_2}
\newcommand{\msqy}{m^2_1}

\newcommand{\incgraph}[2][]{%
  \IfFileExists{pr/#2.pdf}{%
    \includegraphics[#1]{pr/#2.pdf}}{%
    \includegraphics[#1]{#2.pdf}}}

\begin{document}


This is a note on numerical computation of
the decay rate of a false vacuum within a potential
of multiple scalar fields.
The aim is to describe a new strategy to obtain the bounce configuration,
that is based on an earlier idea by Konstandin and Huber
splitting the task into two stages \cite{Konstandin:2006nd}.
The first is to search for an intermediate solution
in one-dimensional spacetime using an improved potential.
In the second, the field profile is stretched smoothly
to that in four dimensions.

One advantage of the original approach is that
the outcome, by construction, must be either
a valid solution of the Euclidean equation of motion or
something clearly wrong.
For instance, this property is not shared by
the improved action method \cite{Kusenko:1995jv}.
The authors of Ref.~\cite{Kusenko:1996jn} use this method
to find a fit and then check its quality
by taking the kinetic to potential ratio within the action.
This ratio must be $-2$ for a field configuration 
to leave the action stationary under an infinitesimal scale transformation
\cite{Derrick:1964ww,Coleman:1977th}.
Being a necessary condition, however,
this is not more than a consistency check and
one cannot be sure that the numerical data
is indeed a bounce until it is put into the equation.

Its virtues notwithstanding, the original approach has drawbacks
due to the use of an improved potential
that render it suboptimal in some circumstances.
Particularly problematic is a case in which
the false and the true vacua are separated far apart.
This can arise for instance
in the Minimal Supersymmetric Standard Model (MSSM)
with a large trilinear term unless it involves a top squark
\cite{Casas:1995pd,Casas:1996de}.
One can put an upper limit on such a term by demanding that
the lifetime of the standard vacuum be longer than the age of the universe
\cite{Endo:2010ya,Hisano:2010re}.
The same criterion has been employed in a study of
maximal flavour violation by a soft trilinear coupling \cite{FVA}.
For this last work,
a modification has been made to the original approach,
which is to be the subject of this article.



\FIGURE{
  \incgraph{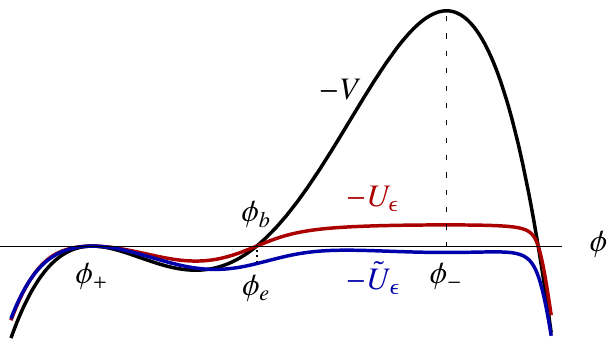}
  \caption{An example potential $V$ 
    and the corresponding improved potentials $U_\epsilon$
    and $\wt{U}_\epsilon$, displayed upside down.
    The false and the true vacua are designated by
    $\phifv$ and $\phitv$, respectively.
    Between them, there are $\phie$ and $\phib$.
    The former denotes any point with $V(\phie) = V(\phifv)$ and
    the latter such a point on the bounce.}
  \label{fig:nondeg}}
Consider the potential depicted in Fig.~\ref{fig:nondeg},
where it has been inverted to make it easy to imagine a classical motion.
It is a function of scalar fields collectively denoted by $\phi$.
The problem is to estimate the rate of transition from
the local minimum $\phifv$ to the state $\phitv$ with the lowest energy.
Note that the notations $\phi_\pm$ in this article follow 
Ref.~\cite{Coleman:1977py}
and are reversed with respect to Ref.~\cite{Konstandin:2006nd}.

Using a semiclassical approximation \cite{Coleman:1977py,Callan:1977pt},
one finds that the decay rate of a false vacuum per unit volume is
\begin{equation}
  \label{eq:probability}
  \Gamma/V = A\,\exp(-\SE[\phibounce]) ,
\end{equation}
where the determinantal factor $A$
comes from the Gaussian functional integral
around the stationary point $\phibounce$, called the bounce,
of the Euclidean action $\SE$.
The formula for $A$ can be found in Ref.~\cite{Callan:1977pt}.
It is very difficult to compute in practice and so
can only be estimated to be around
the characteristic mass scale of the problem.
Normally, evaluation of $A$ is exponentially less important
than that of $\SE[\phibounce]$.
Without loss of generality,
the dominant bounce can be assumed to be $O(4)$-invariant
\cite{Coleman:1977th}.
Therefore, one can let $\phi$ be a function of the single coordinate
$\rho \equiv \sqrt{\tau^2 + \mathbf{x}^2}$,
and write $\SE$ in the form,
\begin{equation}
  \label{eq:SE}
  \SE[\phi(\rho)] = 2\pi^2 \int_0^\infty d\rho\,\rho^3
  \biggl[ \frac{1}{2}\left(\frac{d\phi}{d\rho}\right)^2 + V(\phi) \biggr] .
\end{equation}
The ultimate goal is then to find the solution $\phibounce$ of
the Euler-Lagrange equation,
\begin{equation}
  \label{eq:diff eq}
  \frac{d^2\phi}{d\rho^2} + \frac{3}{\rho} \frac{d\phi}{d\rho} =
  \nabla V(\phi) ,
\end{equation}
derived from the above Euclidean action,
that meets the boundary conditions,
\begin{equation}
  \label{eq:BCs}
  \frac{d\phi}{d\rho} (\rho=0) = 0, \quad
  \phi(\rho \rightarrow \infty) = \phifv .
\end{equation}

The original approach \cite{Konstandin:2006nd} is based on the
idea that one can generalise the equation of motion to
that in $\alpha$ spacetime dimensions,
\begin{equation}
\label{eq:diff eq alpha}
  \frac{d^2\phi}{d\rho^2} +
  \frac{\alpha-1}{\rho} \frac{d\phi}{d\rho} =
  \nabla V(\phi) ,
\end{equation}
which corresponds to the generalised action,
\begin{equation}
  \label{eq:SE alpha}
  \SEalpha[\phi(\rho)] = \int_0^\infty d\rho\,\rho^{(\alpha - 1)}
  \biggl[ \frac{1}{2}\left(\frac{d\phi}{d\rho}\right)^2 + V(\phi) \biggr] .
\end{equation}
The procedure can be divided into two parts:
to solve the equation in the undamped case with $\alpha = 1$
and then to deform the solution to its final shape.
The deformation part has no problem and shall be presented later
when the refined version is described.

The first part for the undamped case is summarised.
If $\alpha = 1$, the damping term in~\eqref{eq:diff eq alpha} disappears
and so the motion conserves energy.
Therefore, one can reproduce the initial condition in~\eqref{eq:BCs}
in the limit where the starting and the ending points are degenerate.
The degeneracy is approximated by flattening out part of the potential
that is lower than the false vacuum.
The solution is found
by minimising the action~\eqref{eq:SE alpha} for $\alpha = 1$.
More specifically, the procedure is made up of the following steps.
\begin{enumerate}
\item
Find the minimum of $\SEx{1}$ with $V$ in~\eqref{eq:SE alpha}
replaced by the improved potential,
\begin{equation}
  \label{eq:Uepsilon}
  U_\epsilon(\phi) = \frac{V(\phi)-V(\phifv)}{2} +
  \left[ \frac{[V(\phi)-V(\phifv)]^2}{4} + \epsilon^2 \right]^{1/2} ,
\end{equation}
satisfying the boundary conditions,
\begin{equation}
  \label{eq:BCs improved potential}
  \phi(0) = \phitv , \quad
  \phi(T) = \phifv .
\end{equation}
The shape of $U_\epsilon$ is illustrated in Fig.~\ref{fig:nondeg}
with $\epsilon$ set to $[V(\phitv) - V(\phifv)]/10$.
One can accelerate the convergence by making
a further modification to $U_\epsilon$ to get
\begin{align}
  \wt{U}_\epsilon(\phi) &= U_\epsilon(\phi) + \Delta U_\epsilon(\phi) ,
\intertext{shown in Fig.~\ref{fig:nondeg}, with a small perturbation,}
  \Delta U_\epsilon(\phi) &=
  - 2 \epsilon \frac{|\phi-\phifv|^3}{|\phitv-\phifv|^3}
  + 3 \epsilon \frac{|\phi-\phifv|^2}{|\phitv-\phifv|^2} .
\label{eq:Delta Uepsilon}
\end{align}
The initial choice of the parameter $\epsilon$ and the time $T$ may be
\begin{align}
\epsilon_\mathrm{init} &= V(\phitv) - V(\phifv) ,
\\
T_\mathrm{init} &= 20\, |\phitv-\phifv| / \sqrt{8 V_b} ,
\label{eq:Tinit}
\end{align}
where $V_b$ is the height of the potential barrier.
\item
Find the point $\phieinit$ between $\phitv$ and $\phifv$
such that $V(\phieinit) = V(\phifv)$
within the configuration obtained in the previous step.
Truncate the part before $\phieinit$, i.e.\ the configuration on the plateau.
\item
Repeat minimisation with the new set of boundary conditions,
\begin{equation}
  \label{eq:BCs float orig}
  \phi(0) = \phie , \quad
  \phi(T) = \phifv ,
\end{equation}
while iteratively sending $\epsilon$ down to zero.
Initially $\phie$ is set to the point found above
and then 
it is allowed to move freely on the submanifold with
$V(\phie) = V(\phifv)$.
\item
The last minimum for $\epsilon = 0$ is the solution of
the undamped equation,
\begin{equation}
  \label{eq:diff eq alpha=1}
  \frac{d^2\phi}{d\rho^2} = \nabla V(\phi) ,
\end{equation}
which is~\eqref{eq:diff eq alpha} for $\alpha = 1$,
and obeys the boundary conditions,
\begin{equation}
  \label{eq:BCs finite T}
  \frac{d\phi}{d\rho} (0) = 0, \quad
  \phi(T) = \phifv.
\end{equation}
\end{enumerate}

There are a couple of points of concern.
First, the improved potential~\eqref{eq:Uepsilon},
sketched in Fig.~\ref{fig:nondeg},
has a plateau on which the solution of step 1 has to spend a long time.
In fact this is what the improved potential is designed for so that
the initial condition in~\eqref{eq:BCs} can be approximated.
As already stated in Ref.~\cite{Konstandin:2006nd},
however, this costs generically many lattice points to store
a long part of the path that is to be discarded eventually.
This grows more and more problematic as $\phitv$ moves away from $\phib$,
and hampers the computation.
Second, step 3 might lead to a numerical instability.
The intention of this step is to find a path that
starts from $\phie$ and then rolls down the inverted potential.
Although modified to have a plateau, 
the improved potential still has a part with
$U_\epsilon(\phi) < U_\epsilon(\phie) = U_\epsilon(\phifv)$
for a finite $\epsilon$.
After starting from $\phie$,
the path may prefer to climb up the inverted potential
and spend time in this region to minimise the action.
This can be avoided by using $\wt{U}_\epsilon$
instead of $U_\epsilon$.
With a judicious choice of $\Delta U_\epsilon$,
one can make $\phifv$ the global minimum of $\wt{U}_\epsilon$
so that the system tries to spend as much time as possible around
$\phifv$ at the end of the path.
Note that for this, $\epsilon$ in~\eqref{eq:Delta Uepsilon}
should be different from $\epsilon$ in~\eqref{eq:Uepsilon}
in general.
The initial condition in~\eqref{eq:BCs finite T} is violated
by $\wt{U}_\epsilon(\phie) - V(\phie)$, which eventually goes away
as $\epsilon \rightarrow 0$.




The first point makes it difficult to analyse many interesting problems.
For instance, vacuum transition triggered by the stau trilinear coupling
in the MSSM is studied in Ref.~\cite{Hisano:2010re}.
The charge-breaking global minimum in the scalar potential is far from
the local one due to the small tau Yukawa coupling.
To overcome this problem,
they first found the bounce
of a temporary potential with two nearly degenerate minima
using the above steps,
and then made a continuation to the actual potential by iteration.
In the course of deforming the potential,
one should be careful not to introduce a singular behaviour.


In what follows, a more streamlined procedure shall be presented.
It does not cost extra lattice points to be truncated in the end.
It does not need a deformation of the potential.
It does not rely on the location of the true vacuum.
Consequently, it can deal with a potential that has
widely separated local and global minima or
that is even unbounded from below.

The main idea is to exploit the energy conservation in the undamped case,
which was already mentioned in Ref.~\cite{Konstandin:2006nd}.
This feature enables one to replace the Neumann boundary condition
of the original problem by a constraint on the potential.
As pointed out above,
fixing the potential at $\rho = 0$ is not enough to
have the desired solution be a minimum since
a path can lower the action further by
shooting up the inverted potential and
staying in a region with $V(\phi) < V(\phifv)$.
The refinement here is to eliminate those paths by additional constraints.

Including the preparation and
the continuation stages as well,
the new series of steps would be as follows.
\begin{enumerate}
\item
Find a point $\phieinit$ such that $V(\phieinit) = V(\phifv)$
that is on the other side of the barrier.
For instance, one can walk over the barrier
along a valley starting from $\phifv$
until the level comes down back to $V(\phifv)$.
\item
Construct an initial configuration $\phi(\rho)$ such that
\begin{equation}
  \phi(0) = \phieinit , \quad
  \phi(T) = \phifv .
\end{equation}
The simplest choice would be a step-like profile
in which $\phi(\rho)$ is constant except for one jump
somewhere in the middle, possibly around $\rho = 0.1 T$.

Choosing a suitable $T$ is important for efficient evaluation.
Recall that \eqref{eq:constraints on ends} can substitute for
\eqref{eq:BCs finite T} only in the limit of energy conservation.
Realising this limit would require infinite $T$, whereas
a small $T$ saves the computation time and storage.
However, it is difficult to estimate an optimal $T$
before doing any minimisation.
Initially, one may try~\eqref{eq:Tinit} with $\phieinit$ instead of
$\phitv$ and $V_b$ from the previous step,
and then $T$ can be adjusted later as explained below.

Optionally to prepare a better initial profile, one could perform
a single-field minimisation of $\SEx{1}$ using as the field
the position on the segment connecting $\phifv$ and $\phieinit$.
One should repeat this process until $T$ is adjusted appropriately.
The criterion for this and the constraints on the potential are
the same as in the next step.
\item
Find a minimum of $\SEx{1}$ obeying the boundary conditions,
\begin{subequations}
  \label{eq:constraints}
  \begin{align}
  \label{eq:constraints on ends}
  V(\phi(0)) &= V(\phifv), \quad
  \phi(T) = \phifv, \\
\intertext{in combination with}
  \label{eq:constraints in between}
  V(\phi(\rho)) &\ge V(\phifv) \quad\text{for}\quad 0 < \rho < T .
  \end{align}
\end{subequations}
The inequality \eqref{eq:constraints in between} should be enforced
at every lattice point in order to prevent the unwanted
`upshooting'.  

One should repeat the minimisation with an increased $T$
until energy is sufficiently conserved.
It usually works to require that
the average difference between the kinetic and the potential energy densities
be less than around 1\% of the barrier height
measured within the resulting profile.
\item
The minimum is a solution of the equation,
\begin{equation}
\label{eq:eq for iteration}
  \frac{d^2\phi}{d\rho^2} +
  \frac{\alpha-1}{\rho + \offset} \frac{d\phi}{d\rho} =
  \nabla V(\phi) ,
\end{equation}
with the damping term killed by $\alpha = 1$, that
complies with the boundary conditions~\eqref{eq:BCs finite T}.
In comparison to~\eqref{eq:diff eq alpha},
there is an additional
offset parameter $\offset$ that has been introduced to avoid a pathological
behaviour near to $\rho = 0$ \cite{Konstandin:2006nd}.
\item
Make a continuation to the damped case by gradually increasing
$\alpha$ from 1 to 4 with $\offset$ fixed around $0.75 T$, and then
send $\offset$ also to zero.
One can choose to stop at $\alpha = 3$ for tunnelling in
a finite temperature system \cite{thermal}.
For each pair of $\alpha$ and $\offset$,
one can linearise~\eqref{eq:eq for iteration}
using the series expansion of the right hand side,
\begin{equation}
  \nabla V(\phi) \approx \nabla V(\wt{\phi}) + (\phi - \wt{\phi}) \cdot
  \nabla \nabla V(\wt{\phi}) ,
\end{equation}
and then iteratively solve it by matrix inversion.
Boundary conditions are set by \eqref{eq:BCs finite T}.
\item
The final solution is the bounce configuration.
\end{enumerate}
It should be reminded that
steps 4 and 5 have been copied from Ref.~\cite{Konstandin:2006nd}.
Modifications have been made only to the steps for the undamped case.

The proof of step 3 follows directly from
that of step 1 in the original method \cite{Konstandin:2006nd}.
First of all, consider a solution determined by
the improved potential~\eqref{eq:Uepsilon}
and the boundary conditions~\eqref{eq:BCs improved potential}.
It is a minimum of the action since 
$\phitv$ is the global minimum of $U_\epsilon$.
Next, take the limit of $\epsilon \rightarrow 0$.
Since the path on the plateau makes no contribution to the action,
the rest of the path becomes a minimum of $\SEx{1}$
subject to the constraints~\eqref{eq:constraints}.
This motion off the plateau satisfies the equation~\eqref{eq:diff eq alpha=1}
and the boundary conditions~\eqref{eq:BCs finite T}
which are equivalent to~\eqref{eq:constraints on ends}
by virtue of energy conservation.
Note that the improved potential is invoked only for
the proof but never appears in the practical numerical calculation.

There is a subtlety regarding the bounce as a minimum of the action
under the conditions~\eqref{eq:constraints}.
In the limit of $T \rightarrow \infty$,
there is a flat direction of minima that initially spend
different amounts of time staying at $\phie$ before rolling down.
Any of this family of minima qualifies in principle as a solution of
the undamped equation.
In practice, however, it is desirable to find a path
that does not have a long constant part to save computation resources.
This is why the initial profile in step 2 was set to have
a jump at a position close to $\rho = 0$.
For a finite $T$, the flat direction is slightly lifted.
The minimum search routine should be able to deal with
this kind of continuously (quasi-)degenerate minima.
Also, the termination condition should be tuned
in such a way that the routine stops after
reaching a point that is acceptable to step 4
even though it is not the exact minimum.
This needs some experiments.

The minimisation problem in step 3 is a nontrivial task
termed nonlinear programming \cite{nlp}
which is a broad research area on its own.
One possible strategy is the following (see e.g.\ \cite{ipopt paper}).
An equality constraint is taken care of by employing a Lagrange multiplier.
An inequality constraint is replaced with an equality
by introducing a slack variable with a limited range.
This limit is fulfilled by adding
a barrier term to the function to be minimised.
This term is a product of the barrier parameter $\mu$
and a barrier function of the slack variable.
The solution of the original problem is obtained by
solving a sequence of subproblems
with $\mu$ successively decreasing towards zero.
At each `outer iteration', the subproblem with the fixed $\mu$
is solved by a damped Newton's method which involves `inner iterations',
and then this and the previous outcomes are used for making
the initial guess for the next subproblem with a smaller $\mu$.
The method in Ref.~\cite{ipopt paper} needs
the `outer loop' to be repeated several times.
If the algorithm is too hard to implement,
one can let a library do the job.
For instance, the example exhibited below uses
the Ipopt package \cite{ipopt}.

It might be of interest to compare this and the original approaches
for a problem that both can solve.  
Suppose that one introduces $n$
lattice points each holding $f$ unfixed field variables
between $\phifv$ and $\phie$ (see Fig.~\ref{fig:nondeg}).
The constrained minimisation problem in step 3 roughly corresponds to
applying Newton's method to
a function of $n (f+2) - 1$ free variables,
i.e.\ $n f$ field values, $n$ Lagrange multipliers, and
$n-1$ slack variables.
This procedure is repeated, each time with a decreased barrier parameter.
To use the original approach on the other hand,
one needs additional $N$ points
from $\phie$ to $\phitv$ except $\phie$ that has already been counted.
The function to be minimised
depends on $(n + N) f$ variables in the first step
and $n f + 1$ in the second including one Lagrange multiplier.
Newton's method is repeated while iteratively driving
$\epsilon \rightarrow 0$.
Given only this information,
it is difficult to tell which way is more efficient on general grounds.
Nevertheless, it is obvious that
a case with large $N$ is better handled with the new strategy
since each Newton iteration becomes very expensive with the original.
This happens when $\phitv$ is distant from $\phie$.
Another limit arises with large $f$
that may favour the new approach over the original.
In this case, the number of auxiliary degrees of freedom $2n-1$
in the former becomes much smaller than $Nf$ in the latter.



Now, the refined procedure is applied to a problem
with two real fields as a demonstration.
The scalar potential is chosen to be
\begin{equation}
  \label{eq:ufb potential}
  V = \frac{\msqy}{2} \y^2 + \frac{\msqx}{2} \x^2
  - \frac{A}{2} \y\x^2 + g^2 (\y^2 - \x^2)^2 .
\end{equation}
This potential is unbounded from below.
In this sense, it is qualitatively
different from the examples in Ref.~\cite{Konstandin:2006nd}.
Clearly, one cannot impose
the boundary conditions~\eqref{eq:BCs improved potential}
since there exists no $\phitv$.
Therefore, the original method is not applicable unless
the potential is altered.
There is a minimum with $V = 0$ at $\y = \x = 0$ that is local
as long as $A \ne 0$.
One can see the metastability easily by taking the field direction,
$\y = \x = a$, that cuts out the quartic term leaving
\begin{equation}
  \label{eq:flat direction}
  V_\mathrm{flat} = \frac{1}{2} (\msqy + \msqx)\,a^2 - \frac{A}{2}\,a^3 .
\end{equation}
Along this direction, a barrier starts from the origin
and ends at the position where $A\,a = \msqy + \msqx$,
after which the potential drops below zero and keeps falling down.

\FIGURE[t]{
\incgraph{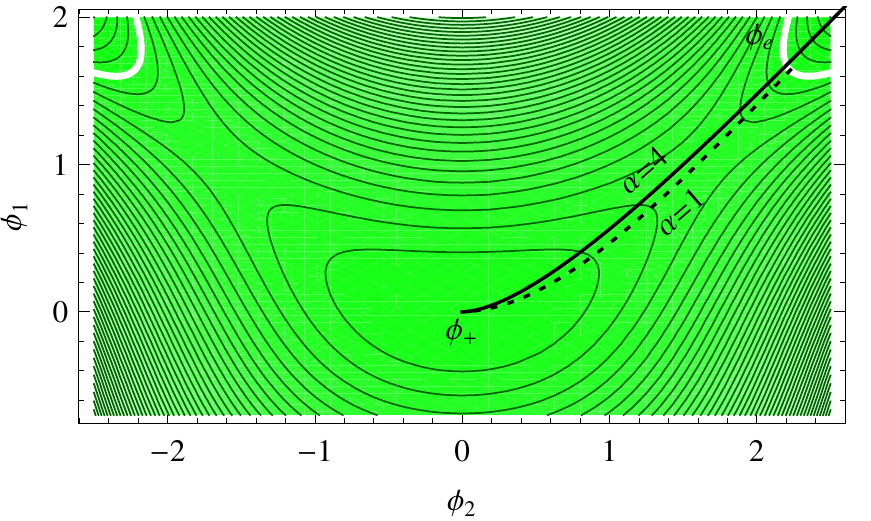}
\caption{Contours of the scalar potential~\eqref{eq:ufb potential}.
  The local minimum is marked with $\phifv$.
  The white curve is the set of points $\phie$ at the same level as $\phifv$.
  The dotted and the solid curves show
  the bounces in the undamped and the damped cases, respectively.}
\label{fig:contours}}
The shape of the potential is illustrated in Fig.~\ref{fig:contours}.
The masses and couplings in~\eqref{eq:ufb potential} are set to be
\begin{equation}
  \label{eq:parameters}
  \msqy = 3.0, \quad
  \msqx = 0.5, \quad
  A = 1.5, \quad
  g = 0.4 .
\end{equation}
There is a force that drives the path off
the straight directions with $\y = \pm\x$,
arising from the gradients of both the trilinear and the mass terms.
A large difference between $\msqy$ and $\msqx$ has been introduced
to emphasise this dynamics.

The white contour corresponds to the point $\phie$ in Fig.~\ref{fig:nondeg}.
Finding any point on the curve is enough to complete step 1.
Successful execution up to step 3 should lead to the
undamped solution for $\alpha = 1$ shown in Fig.~\ref{fig:contours}.
Notice that the starting point of
the trajectory is on the white curve
as enforced by~\eqref{eq:constraints on ends}.
Then one can proceed to take steps 4 and 5 to
deform this path to the
final bounce for $\alpha = 4, \offset = 0$.
It is also partly visible in the same figure.

\FIGURE{
\incgraph{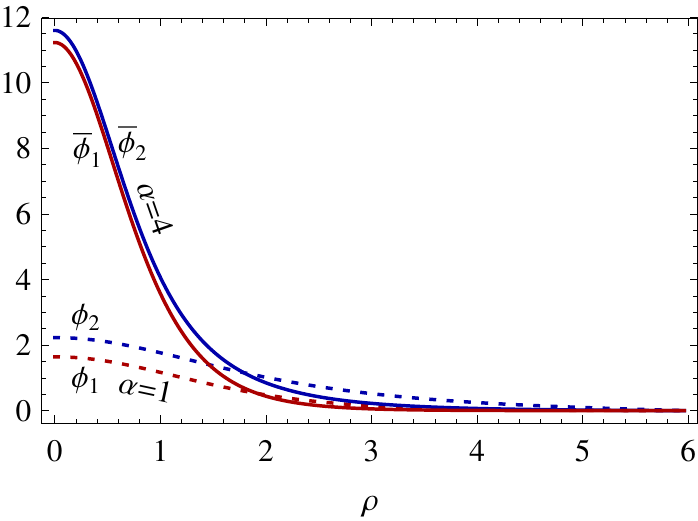}
\caption{Field profiles of the bounces.
  The dotted and the solid curves show
  the undamped and the damped cases, respectively.}
\label{fig:sol}}
The solutions are plotted as functions of the time in Fig.~\ref{fig:sol}.
As $\alpha$ increases,
the starting point retreats from the false vacuum
due to the damping term in~\eqref{eq:eq for iteration}.
Notice that the path does not diverge indefinitely even with an
unbounded-from-below potential.
This is understandable from the fact that
the original problem formulated in~\eqref{eq:diff eq}
and~\eqref{eq:BCs} has nothing to do with
any information on the global minimum.
For the same reason,
the procedure described in this article
does not need to explore deep into the potential
for an unreachable bottom.
One can put this solution back into the Euclidean action~\eqref{eq:SE}
to calculate the final answer,
\begin{equation}
  \SE[\phibounce] = 488 .
\end{equation}
In this problem, the potential has a symmetry with respect to the reflection
$\x \rightarrow -\x$.
This pairs a bounce with the other that makes exactly the same
contribution to the decay rate.
Therefore, one must multiply the decay rate~\eqref{eq:probability} by 2.

It is straightforward to write the
latticised versions of the fields, the Euclidean action,
the Euler-Lagrange equation, and the boundary conditions,
and they can be found in Ref.~\cite{Konstandin:2006nd} for instance.
The same reference reports an analysis of discretisation error.
It is determined by the final iteration
step for continuation to four dimensions
that is common to the present and the original methods.


\acknowledgments

The author thanks Ahmed Ali for helpful comments.


\end{document}